\documentclass[10pt,a4paper]{article}
\usepackage{a4wide}
\usepackage{latexsym}
\usepackage{amssymb}
\usepackage{amsmath, cite}
\usepackage{color}

 \makeatletter
\@addtoreset{equation}{section} \makeatother

\def\bfone{\relax{\rm 1\kern-.35em 1}}

 \makeatletter
\@addtoreset{equation}{section} \makeatother

\def\be {\begin{equation}}
\def\ee {\end{equation}}
\def\bea {\begin{eqnarray}}
\def\eea {\end{eqnarray}}
\def\bc {\begin{center}}
\def\ec {\end{center}}
\def\a  {\alpha}
\def\b  {\beta}

\def\D  {\Delta}

\def\bfg {\begin{figure}}
\def\efg {\end{figure}}
\def\bi {\begin{itemize}}
\def\ei {\end{itemize}}
\def\nn {\nonumber}

\def\la {\label}
\def\le {\left}
\def\ri {\right}

\def\fr {\frac}

\DeclareFontFamily{U}{rsf}{} \DeclareFontShape{U}{rsf}{m}{n}{
  <5> <6> rsfs5 <7> <8> <9> rsfs7 <10-> rsfs10}{}
\DeclareMathAlphabet\Scr{U}{rsf}{m}{n}

\begin{document}

\begin{center}
{\bf \large{ Planck-Scale Corrections to Friedmann Equation  \\[10mm]}}
\large Adel Awad$^{
\star \dagger \ddagger}$\footnote{Electronic address~:~\upshape{aawad@zewailcity.edu.eg }} and  Ahmed Farag Ali$^{\star \S}$\footnote{\upshape{Electronic address~:~ahmed.ali@fsc.bu.edu.eg~;~afarag@zewailcity.edu.eg}}
\\[7mm]
\small{$^\star$Centre for Fundamental Physics, Zewail City of Science and Technology}\\ \small{ Sheikh Zayed, 12588, Giza, Egypt.}\\[4mm]
\small{$^\dagger$Department of Physics, Faculty of Science, Ain Shams University, Cairo, 11566, Egypt}\\[4mm]
\small{$^\dagger$Center for Theoretical Physics, British University of Egypt, Sherouk City 11837, P.O. Box 43, EGYPT}\\[4mm]
\small{$^{\S}$Department of Physics, Faculty of Science, Benha University, Benha 13518, Egypt.}
\end{center}

\begin{abstract}

Recently, Verlinde proposed that gravity is an emergent phenomenon which originates from an entropic force.
In this work, we extend Verlinde's proposal to accommodate generalized uncertainty principles (GUP), which are suggested by some approaches to \emph{quantum gravity} such as string theory, black hole physics and doubly special relativity (DSR). Using Verlinde's proposal and two known models of GUPs, we obtain modifications to Newton's law of gravitation as well as the Friedmann equation. Our modification to the Friedmann equation includes higher powers of the Hubble parameter which is used to obtain a corresponding Raychaudhuri equation. Solving this equation, we obtain a leading Planck-scale correction to Friedmann-Robertson-Walker (FRW) solutions for the $p=\omega \rho$ equation of state.
\end{abstract}




\section{Introduction}

One of the intriguing predictions of various frame works of quantum gravity,
such as string theory and black hole physics, is
the existence of a minimum measurable length.~This has given rise to the so-called generalized uncertainty principle (GUP) or equivalently, modified commutation relations between position coordinates
and momenta \cite{guppapers,BHGUP0,BHGUP1,BHGUP2,BHGUP3,BHGUP4,BHGUP5,Scardigli}.~This can be understood in the context of string theory since strings can not
interact at distances smaller than their size.~The GUP is represented by the following form
\cite{kmm,kempf,brau}:
\bea \Delta x_i \Delta p_i &\geq& \fr{\hbar}{2} [ 1 + \beta
\le((\Delta p)^2 + <p>^2 \ri)
+ 2\beta \le( \Delta p_i^2 + <p_i>^2\ri) ]~, \la{uncert1}
\eea
where $p^2 = \sum\limits_{j}p_{j}p_{j}$,
$\beta=\beta_0/(M_{p}c)^2=\b_0 \frac{\ell_{p}^2}{\hbar^2}$, $M_{p}=$
Planck mass, and $M_{p} c^2=$ Planck energy.
The inequality (\ref{uncert1}) is equivalent to the following modified Heisenberg algebra \cite{kmm}:
\be [x_i,p_j] = i \hbar ( \delta_{ij} + \beta \delta_{ij} p^2 +
2\beta p_i p_j )~. \la{com1} \ee
This form ensures, via the Jacobi identity, that
$[x_i,x_j]=0=[p_i,p_j]$ \cite{kempf}.\\

Recently, a new form of GUP was proposed in
\cite{advplb,Das:2010zf},~which predicts maximum observable momenta as well as the existence of minimal measurable
length, and is consistent with doubly special relativity theories (DSR)\cite{cg}, string theory and black holes physics \cite{guppapers,BHGUP0,BHGUP1,BHGUP2,BHGUP3,BHGUP4,BHGUP5}.~It
ensures $[x_i,x_j]=0=[p_i,p_j]$, via the Jacobi identity.
\bea [x_i, p_j]\hspace{-1ex} &=&\hspace{-1ex} i
\hbar\hspace{-0.5ex} \left[  \delta_{ij}\hspace{-0.5ex} -
\hspace{-0.5ex} \alpha \hspace{-0.5ex}  \le( p \delta_{ij} +
\frac{p_i p_j}{p} \ri)
+ \alpha^2 \hspace{-0.5ex}
\le( p^2 \delta_{ij}  + 3 p_{i} p_{j} \ri) \hspace{-0.5ex} \ri],
\label{comm01}
\eea
where $\alpha = {\alpha_0}/{M_{p}c} = {\alpha_0 \ell_{p}}/{\hbar},$
$M_{p}=$ Planck mass, $\ell_{p}=$ Planck length,
and $M_{p} c^2=$ Planck energy.
In a series of papers, various applications of the new model of GUP have been investigated \cite{dvprl,dvcjp, Ali:2011fa,faragali,neutrinoGUP,Chemissany:2011nq,Ali:2012mt0,Ali:2012mt1,Tawfik:2012he,
Majumder:2012qy,Nozari:2012nf,Nozari:2012gd,Majumder:2012ph}.

The upper bounds on the GUP parameter $\alpha$ has been derived in \cite{Ali:2011fa}.
It was suggested that these bounds can
be measured using quantum optics and gravitational wave techniques
in \cite{Pikovski:2011zk,NatureGRW}. Recently, Bekenstein \cite{Bekenstein:2012yy,Bekenstein:2013ih} proposed that quantum gravitational effects could be tested experimentally, suggesting``a tabletop experiment which,
given state of the art ultrahigh vacuum and cryogenic technology,
could already be sensitive enough to detect Planck scale signals'' \cite{Bekenstein:2012yy}.
This would enable several quantum gravity predictions to be tested in the Laboratory \cite{Pikovski:2011zk,NatureGRW}. This is considered as a milestone in the field of quantum gravity phenomenology.

Motivated by the above arguments, we investigate possible effects of GUP on
the Friedmann equation.~We use the entropic force approach suggested by Verlinde \cite{everlind1} to calculate corrections
to Newton's law of gravitation and the Friedmann equations for two models of GUP, which are mentioned above. We found that Planck-scale corrections to the Friedmann equation include higher powers of the Hubble parameter which are suppressed by Planck length. Using these corrections we construct the corresponding Raychaudhuri equations, which we then solve to obtain a leading Planck-scale correction to the Friedmann-Robertson-Walker (FRW) solutions with equation of state $p=\omega \rho$. Deriving the effects of GUPs on Newton's law of gravitation through the entropic force approach was initially reported in \cite{Ali:2013ma} where a modified Newton's law of gravitation was calculated. A possible application of these Planck-scale corrections is early cosmology, in particular inflation models where physical scales are only a few orders of magnitude less than Planck scale.

Let us review Verlinde's proposal \cite{everlind1} on the origin of gravity, where he suggested that
a gravitational force might be of an entropic nature.
This assumption is based on the relation
between gravity and thermodynamics \cite{bhtherm0,bhtherm1,bhtherm2}.
According to thermodynamics and the holographic principle,
Verlinde's approach results in Newton's law of gravitation, the Einstein Equations \cite{everlind1}
and the Friedmann Equations\cite{Cai:2010hk}.
The theory can be summarized as \\
at the temperature $T$, the entropic force $F$ of  a gravitational system is given as:
\be
F \Delta x = T \Delta S, \la{force}
\ee
where $\Delta S$ is the change in the entropy such
that, at a displacement $\Delta x$,
each particle carries its own portion of entropy change.
From  the correspondence between the entropy change
$\Delta S$ and the information
about the boundary of the system and using Bekenstein's
argument \cite{bhtherm0,bhtherm1,bhtherm2}, it is assumed that $ \Delta S = 2 \pi k_B$,
where $\Delta x = \hbar/m$ and $k_B$ is the Boltzmann constant.
\be
\Delta S = 2 \pi k_B \frac{m c}{\hbar} \Delta x, \label{entropy}
\ee
where $m$ is the mass of the elementary component, $c$ is speed
of light and $\hbar$ is the Planck constant.

The holographic principle assumes that for a region enclosed by some surface gravity
can be represented by the degrees of freedom on the surface
itself and independent of the its bulk geometry. This implies
that the quantum gravity can be described by a topological
quantum field theory, for which all physical degrees of freedom
can be projected onto the boundary\cite{thooft}. The information
about the holographic system is given by  $N$ bits forming an ideal gas.
It is conjectured that $N$ is proportional to the entropy of the holographic screen,

\bea
N= \frac{4 S}{k_B},\la{Nbits}
\eea
then according to Bekenstein's entropy-area relation \cite{bhtherm0,bhtherm1,bhtherm2}
\be
S=\frac{k_B c^3}{4 G \hbar} A.
\ee
Therefore, one gets
\be
N=\frac{A c^3}{G \hbar} = \frac{4 \pi r^2 c^3}{G \hbar},
\ee
where $r$ is the radius of the gravitational system and $A= 4 \pi r^2$
is area  of the holographic screen.
It is assumed that each bit emerges out of the
holographic screen i.e. in one dimension. Therefore each bit
carries an energy equal to $k_B T/2$. Using the equipartition
rule to calculate the energy of the system, one gets
\be
E=\frac{1}{2} N k_B T = \frac{2 \pi c^3 r^2}{G \hbar} k_B T = M c^2. \la{energy}
\ee
By substituting Eq. (\ref{force}) and Eq. (\ref{entropy}) into Eq. (\ref{energy}), we get
\be
F=G \frac{M m}{r^2},
\ee
making it clear that Newton's law of gravitation can be derived
.\\

Recently, a modified
Newton's law of gravity due to Planck-scale effects through the entropic force approach was derived by one of the authors \cite{Ali:2013ma}.
Derivation of Planck scale effects on the Newton's
law of gravity are based on the following procedure:
\emph{modified theory of gravity $\rightarrow$ modified black hole entropy$\rightarrow$
modified holographic surface entropy $\rightarrow$ Newton's law corrections}. This procedure has been
followed with other approaches like non-commutative geometry in \cite{Nicolini:2010nb,Bastos:2010au,Nozari:2011et,Mehdipour:2012nj}.
In this paper, we take into account
the quantum gravity corrections due to GUP in the entropic-force approach, following the same
procedure as \cite{Ali:2013ma}, and extend our study to calculate the modified Friedmann and Raychaudhuri equations.
To calculate the quantum gravity corrections to the Friedmann equations, we use the procedure that has been followed in
\cite{Sheykhi:2010wm,Hendi:2010xr}. It is worth mentioning that recently,
there have been some considerable interest in entropic force approach and its applications  \cite{application}.

An outline of this paper is as follows. In Sec .2, We
investigate entropic force if the GUP of Eq.\
(\ref{uncert1}), which was proposed earlier by
\cite{guppapers,Scardigli,BHGUP0,BHGUP1,BHGUP2,BHGUP3,BHGUP4,BHGUP5,kmm,kempf,brau}, is taken into
consideration. First we calculate the modified thermodynamics
of the black hole which yields a modified entropy. By the holographic principle,
we calculate the modified number of bits $N$. Based on the modified number of bits,
we estimate the Planck scale correction to
Newton's law of gravitation and the Friedmann equations. We solve Raychaudhuri equations
to obtain Planck scale corrections to FRW cosmological solutions.

In Sec. 3, we repeat the analysis for the newly proposed
model of GUP  in Eq.\ (\ref{comm01}).  We calculate different
corrections to Newton's law of gravity and the Friedman equations.
We compare our results with previous studies of quantum gravity corrections
to gravity laws. We then solve the modified Raychaudhuri equations.
Further implications are discussed.

\section{GUP-quadratic in $\D p$ and BH thermodynamics}
\label{sec:GUP-QUAD}

In this section, we review  the modified thermodynamics
of the black hole which yields a modified entropy due to GUP \cite{Eliasentropy0,Eliasentropy1,Adler,Cavaglia:2003qk0,Cavaglia:2003qk1,Cavaglia:2003qk2,Scardigli:1995qd}. Using the holographic principle,
we get a modified number of bits $N$ which yields quantum gravity corrections to
Newton's law of gravitation and the Friedmann equations.

Black holes are considered as a good laboratories
for the clear connection between thermodynamics and gravity, so
black hole thermodynamics will be analyzed in this section.
We then make an analysis of BH thermodynamics if the GUP-quadratic
in $\D p$ that was proposed in \cite{guppapers,Scardigli,BHGUP0,BHGUP1,BHGUP2,BHGUP3,BHGUP4,BHGUP5,kmm,kempf,brau}
is taken into consideration.
With Hawking radiation, the emitted particles are mostly
photons and standard model (SM) particles.
Using the Hawking-Uncertainty Relation, the characteristic energy of the emitted
particle can be identified  \cite{Adler,Cavaglia:2003qk0,Cavaglia:2003qk1,Cavaglia:2003qk2,Scardigli:1995qd}.
It has been found \cite{Ali:2012mt0,Ali:2012mt1}, assuming some symmetry conditions
from the propagation of Hawking radiation, that the inequality
that would correspond to Eq. (\ref{uncert1}) can be written as follows:

\be \D x \D p \geq \frac{\hbar}{2} \le[1+ \frac{5}{3}~
(1+\mu)~\b_0~ \ell_p^2~ \frac{\D p^2}{\hbar^2} \ri]\,.
\la{ineqI} \ee

where $\mu= \le( \frac{2.821}{\pi}\ri)^2$.

By solving the inequality (\ref{ineqI}) as a quadratic equation
in $\D p$, we obtain

\be \frac{\D p}{\hbar}\geq\frac{\D x}{\frac{5}{3}~(1+\mu)~\b_0
\ell_{p}^2}\le[1- \sqrt{1-\frac{\frac{5}{3}~(1+\mu)~
\b_0\ell_{p}^2} { \D x^2}}~~\ri] \,.\la{gupsol} \ee

Using Taylor expansion, Eq. (\ref{gupsol}) reads

\be
\Delta p \geq \frac{1}{2 \D x}\le[1+ \frac{5}{12} \frac{(1+\mu) \b_0 \ell_p^2}{\D x^2} + \mathcal{O}(\b_0^2)\ri],
\ee

where we used natural units as $\hbar=1$. According to \cite{Eliasentropy0,Eliasentropy1} a photon is used to ascertain the position of a quantum particle of energy $E$
and according to the argument in \cite{AmelinoCamelia:2004xx0,AmelinoCamelia:2004xx1} which states that the uncertainty principle
$\D p \geq 1/ \D x$  can be translated to the lower bound $E\geq 1/ \D x$, one can write for the GUP case

\be
E \geq \frac{1}{2 \D x}\le[1+ \frac{5}{12} \frac{(1+\mu) \b_0 \ell_p^2}{\D x^2} + \mathcal{O}(\b_0^2)\ri] \la{energy0}
\ee

For a black hole absorbing a quantum particle with energy $E$ and size $R$, the area of the black hole should increase by \cite{Areachange}.
\be
\Delta A \geq 8 \pi\, \ell_p^2\, E\, R,
\ee
The quantum particle itself implies the existence of finite bound given by
\be
\Delta A_{min} \geq 8 \pi\, \ell_p^2\, E\, \Delta\, x. \la{Darea}
\ee
Using the Eq. (\ref{energy0}) in the inequality (\ref{Darea}), we get
\be
\Delta A_{min} \geq 4 \pi\, \ell_p^2 \le[1+ \frac{5}{12} \frac{(1+\mu) \b_0 \ell_p^2}{\D x^2} + \mathcal{O}(\b_0^2)\ri]. \la{Area}
\ee

The value of $\Delta x$ is taken to be inverse of surface gravity $\D x= \kappa^{-1}= 2 r_s$ where $r_s$ is the
Schwarzschild radius. Where this is  probably the most sensible choice of length scale is
in the context of near-horizon geometry \cite{Eliasentropy0,Eliasentropy1}.
This implies that
\be
\D x^2 = \frac{A}{\pi} \label{DX}.
\ee

Substituting  Eq. (\ref{DX}) into Eq. (\ref{Area}), we get

\be
\Delta A_{min} \simeq \lambda \ell_p^2 \le[1+ \frac{5 ~\pi }{12} ~\frac{(1+\mu)~ \b_0 \ell_p^2}{A} + \mathcal{O}(\b_0^2)\ri]. \la{Area}
\ee

where the parameter $\lambda$  will be defined later. According to \cite{bhtherm0,bhtherm1,bhtherm2}, the black hole's entropy is conjectured to depend on the horizon's area. From information theory \cite{Adami:2004mx}, it has been found that the smallest increase in entropy should be independent of the area. It is just
one bit of information, which is $b = \ln(2)$.

\be
\frac{dS}{dA}= \frac{\Delta S_{min}}{\Delta A_{min}} = \frac{b}{\lambda \ell_p^2 \le[1+ \frac{5 ~\pi }{12} ~\frac{(1+\mu)~ \b_0 \ell_p^2}{A} + \mathcal{O}(\b_0^2)\ri]},
\ee

where $b$ is a parameter. By expanding the last expression in orders of $\b_0$ and then integrating, we get the entropy

\be
S= \le[\frac{b}{\lambda}\frac{A}{\ell_p^2}- \frac{5 ~\pi~(1+\mu)}{12} \b_0 \frac{b}{\lambda} ~\ln{\le( \frac{A}{\ell_p^2}\ri)} \ri].
\ee

Using the Hawking-Bekenstein assumption \cite{bhtherm0,bhtherm1,bhtherm2}, which relates entropy with the area, the value of $b/ \lambda$ is fixed to be~$ = 1/4$, so that

\be
S=\le(\frac{A}{4\, \ell_p^2}\ri) -  \frac{5 ~\pi~(1+\mu)}{48} \b_0  ~\ln{\le(\frac{A}{ \ell_p^2}\ri)}. \la{correctENTROPY0}
\ee

It is concluded that the entropy is directly related to the area and gets \emph{a logarithmic correction} when applying GUP approach\cite{Eliasentropy0,Eliasentropy1}.

The temperature of the black hole is

\bea
T&=& \frac{\kappa}{8 \pi} \frac{dA}{dS}
=\frac{\kappa}{8 \pi} \le[1+ \frac{5 ~\pi }{48} ~\frac{(1+\mu)~ \b_0 \ell_p^2}{A} + \mathcal{O}(\b_0^2)\ri].
\eea
So far, it is concluded that the temperature is not only proportional to the surface gravity but
also it depends on the black hole's area\cite{Eliasentropy0,Eliasentropy1}.

\subsection{Modified Newton's law of gravity}

In this section we   study the implications of the corrections
for the entropy in  Eq.~(\ref{correctENTROPY})
and calculate how the number of bits of Eq.~(\ref{Nbits})
would be modified leading to a new correction
to Newton's law of gravitation. Using the corrected entropy
given in Eq. (\ref{correctENTROPY0}), we find that the number of bits
should be corrected as follows:

\be
N^{\prime}= \frac{4 S}{k_B}= \le(\frac{A}{ \ell_p^2}\ri) -  \frac{5 ~\pi~(1+\mu)}{12} \b_0  ~\ln{\le(\frac{A}{ \ell_p^2}\ri)}. \la{bits0}
\ee
We can substitute for the Planck length with $\ell_p= \frac{\sqrt{\hbar ~G}}{c^{3/2}}$, then we get the
modified number of bits as follows
\be
N^{\prime}= \frac{4 S}{k_B}= \le(\frac{A~ c^3}{ \hbar ~G}\ri) -  \frac{5 ~\pi~(1+\mu)}{12} \b_0  ~\ln{\le(\frac{A~ c^3}{\hbar ~G}\ri)}. \la{bits00}
\ee
We define the constant $\delta=\frac{5 ~\pi~(1+\mu)}{12} \b_0$, then we have
\be
N^{\prime}= \frac{4 S}{k_B}= \le(\frac{A~ c^3}{ \hbar ~G}\ri) -  \delta  ~\ln{\le(\frac{A~ c^3}{\hbar ~G}\ri)}. \la{bits001}
\ee
By substituting Eq. (\ref{bits001}) into Eq. (\ref{energy}) and using Eq. (\ref{force}), we get
\be
E= M c^2= \frac{F ~r^2~ c^2}{m~G} \le(1- \delta ~ \frac{\ln{ (\frac{4 \pi c^3}{\hbar G}r^2)}}{ ~ \frac{2 \pi ~c^3}{\hbar G} r^2}\ri).
\ee
This implies a modified Newton's Law of gravity   given as
\be
F= \frac{G M m}{r^2}\le(1+  \delta ~ \frac{\ln{ (\frac{4 \pi r^2}{\ell_p^2})}}{ 4 \pi r^2/\ell_p^2}\ri). \la{GravForce}
\ee
Which means that the Newtonian gravitational potential would be:
\be
V(r)= -\frac{G M m}{r} \le(1+\frac{2}{9} \delta \frac{1}{4 \pi r^2/\ell_p^2} +\frac{1}{3} \delta
\frac{\ln{ (\frac{4 \pi r^2}{\ell_p^2})}}{ 4 \pi r^2/\ell_p^2}\ri)
\ee
We note that these logarithmic corrections to the Newtonian gravitational potential have
been obtained using an independent approach studying the brane
world corrections to Newton's law of gravity \cite{Bronnikov:2006jy} which
suggest that brane world and GUP would predict similar physics.
In the next subsection, we  discuss modification of the Friedmann equation due
to the corrections of quantum gravity.

\subsection{Modified Friedmann and Raychaudhuri Equations}
\label{sec:GUP-Friedmann}

In this section, we use the analysis of \cite{Sheykhi:2010wm,Hendi:2010xr,Cai:2010hk,app-horizon} which derived the Friedmann
equations using an entropic force approach. We
investigate the effect of the quantum gravity arising from the GUP
proposed in Eq. (\ref{uncert1}) on the form of the Friedmann equations using the method of\cite{Cai:2010hk}.
The Friedmann-Robertson-Walker (FRW) universe is described by the following metric:
\be
ds^2={h}_{a b}dx^{a} dx^{b}+\tilde{r}^2 d\Omega^2, \label{metric}
\ee
where $\tilde{r}= a(t) r,~ x^a=(t,r), h_{a b}=(-1, a^2/(1-kr^2)),d\Omega^2=d\theta^2+\sin^2\theta
d\phi^2 $ and $a,b=0,1$. $k$ is the spatial curvature and it takes the values $0,1,-1$ for
a flat, closed and open universe, respectively.
The dynamic apparent horizon is determined by the relation
$h^{ab}\partial_{a}\tilde{r}\partial_{b}\tilde{r}=0$, which yields its radius\cite{app-horizon}:
\be
\la{app-radius}
 \tilde{r}=a ~r=\frac{1}{\sqrt{H^2+k/a^2}},
\ee
where $H=\dot{a}/a$ is the Hubble parameter.
By assuming that the matter which occupies the FRW universe forms a perfect fluid, the energy-momentum
tensor would be:
\be
T_{\mu\nu}=(\rho+p)u_{\mu}u_{\nu}+pg_{\mu\nu}.\la{stress-tensor}
\ee
The energy conservation law then leads to the continuity equation
\be
\dot{\rho}+3H(\rho+p)=0,\la{Continuity}
\ee
To calculate the quantum gravity corrections to the Friedmann equations, we consider
a compact spatial region with volume $V$ and radius $\tilde{r}= a(t) r$ \cite{Cai:2010hk}.
By combining  Newton's second law for the test particle $m$ near the surface
with the modified gravitational force of Eq. (\ref{GravForce}), we get
\be
\label{F6}
F=m~\ddot{\tilde{r}}=m~\ddot{a}(t)~r=-\frac{GMm}{\tilde{r}^2}\le(1+  \delta ~ \frac{\ln{ (\frac{4 \pi \tilde{r}^2}{\ell_p^2})}}{ 4 \pi \tilde{r}^2/\ell_p^2}\ri).
\end{equation}
Equation (\ref{F6}) can be written in terms of the matter density $\rho= M/ V, V= (4/3) \pi \tilde{r}^3$ as:
\be
\frac{\ddot{a}}{a}=-\frac{4\pi
G}{3}\rho\le(1+  \delta ~ \frac{\ln{ (\frac{4 \pi \tilde{r}^2}{\ell_p^2})}}{ 4 \pi \tilde{r}^2/\ell_p^2}\ri), \la{New-accel}
\ee
The last equation is considered to be the modified Newtonian cosmology.
To derive the modified Friedmann equations, it was assumed as in \cite{Cai:2010hk}, that
the active gravitational mass $\mathcal{M}$, where
\be
\mathcal{M}= 2
\int_V{dV\left(T_{\mu\nu}-\frac{1}{2}Tg_{\mu\nu}\right)u^{\mu}u^{\nu}},\label{activeM}
\ee
generates the acceleration rather than the total mass $M$ in the volume $V$.

So for the FRW universe, the active gravitational mass would be
\be
\mathcal M =(\rho+3p)\frac{4\pi}{3}a^3 r^3.\label{activeM2}
\ee
By replacing M with $\mathcal{M}$,  Eq. (\ref{New-accel}) can be rewritten as:
\be
\mathcal{M}=-\frac{\ddot{a}
a^2}{G}r^3\le(1+  \delta ~ \frac{\ln{ (\frac{4 \pi \tilde{r}^2}{\ell_p^2})}}{ 4 \pi \tilde{r}^2/\ell_p^2}\ri)^{-1}.\label{activeM3}
\ee
By Equating Eq.(\ref{activeM2}) with Eq. (\ref{activeM3}), we get
\be
\frac{\ddot{a}}{a} =-\frac{4\pi
G}{3}(\rho+3p)\le(1+  \delta ~ \frac{\ln{ (\frac{4 \pi \tilde{r}^2}{\ell_p^2})}}{ 4 \pi \tilde{r}^2/\ell_p^2}\ri).\label{acceleration3}
\ee
Using the continuity equation of Eq.(\ref{Continuity}) in Eq.(\ref{acceleration3}), multiplying both sides with $(a~\dot{a})$and integrating, we obtain the following equations:
\be
\dot{a}\ddot{a}= -\frac{4\pi G}{3} \dot{a} a \le(\rho - \frac{\dot{\rho}}{H}- 3 \rho\ri)
\le(1+  \delta ~ \frac{\ln{ (\frac{4 \pi \tilde{r}^2}{\ell_p^2})}}{ 4 \pi \tilde{r}^2/\ell_p^2}\ri).
\ee
%
\be
\frac{d}{dt} (\dot{a}^2) = \frac{8\pi G}{3} \le(\frac{d}{dt} \le(\rho a^2\ri)\ri)
\le(1+ \delta ~ \frac{\ln{ (\frac{4 \pi \tilde{r}^2}{\ell_p^2})}}{ 4 \pi \tilde{r}^2/\ell_p^2}\ri).
\ee
Integrating both sides for each term of the last equation yields
%
\be
\dot{a}^2+k= \frac{8 \pi G}{3} \rho a^2 \le(1+ \frac{\delta}{\rho a^2}\int \frac{\ln{ (\frac{4 \pi a^2 r^2}{\ell_p^2})}}{ 4 \pi a^2 r^2 /\ell_p^2} d\le(\rho a^2\ri)\ri).
\ee
We can take $r$ outside the integration because it does not depend on $a$  from the definition of
$\tilde{r}= r a(t)$ after Eq.(\ref{metric}), we get
\be
\dot{a}^2+k=\frac{8 \pi G}{3} \rho a^2\le[ 1+ \frac{\delta~ \ell_p^2}{4 \pi \rho a^2 r^2} \int \ln{ \le(\frac{4 \pi a^2 r^2}{\ell_p^2}\ri)} \frac{d(\rho a^2)}{a^2} \ri].\la{adot}
\ee
Suppose that the  equation of state parameter $w=p/\rho$ is a constant of time, so the  continuity equation
(\ref{Continuity}) is integrated to give
\begin{equation}\label{rho}
\rho=\rho_0 a^{-3(1+w)},
\end{equation}
By substituting Eq.(\ref{rho}) into Eq.(\ref{adot}) and solving the integration, we get
%
%
\be
H^2+\frac{k}{a^2}=\frac{8 \pi G}{3}\rho\le(1+\frac{2 \delta~ \rho_0 \ell_p^2 (-1-3\omega)}{4 \pi \rho a^2 r^2} \int
 \ln{ \le(\frac{2 \sqrt{\pi}  r}{\ell_p} a \ri)} \le(a^{-4-3\omega}\ri) da  \ri).
\ee
The integration in the last equation can be solved to yield
\be
H^2+\frac{k}{a^2}= \frac{8 \pi G}{3} \rho\le[1+\frac{\delta~ \ell_p^2~(1+3 \omega)}{18 \pi \tilde{r}^2 (1+\omega)^2}
\le(1+3(1+\omega) \ln{\frac{2\sqrt{\pi}~r~a}{\ell_p}}\ri)\ri].
\ee
Using the relation  $\tilde{r}=a ~r=\frac{1}{\sqrt{H^2+k/a^2}}$, we get
\be
\le(H^2+\frac{k}{a^2}\ri)\le[1-\frac{\delta \ell_p^2(1+3\omega)}{18 \pi (1+\omega)^2}(H^2+\frac{k}{a^2})
\le(1-\frac{3(1+\omega)}{2}\ln{\le(H^2+\frac{k}{a^2}\ri)\frac{\ell_p^2}{4 \pi}}\ri)\ri] = \frac{8 \pi G}{3}\rho,
\ee
or it can be written as
\bea
&&\le(H^2+\frac{k}{a^2}\ri)\le[1-\frac{\delta~ \ell_p^2(1+3\omega)}{18 \pi (1+\omega)^2}(H^2+\frac{k}{a^2})
+\frac{\delta~ \ell_p^2(1+3\omega)}{12 \pi(1+\omega) }(H^2+\frac{k}{a^2})\ln\le({(H^2+\frac{k}{a^2})\frac{\ell_p^2}{4 \pi}}\ri)\ri]\nn\\ &&= \frac{8 \pi G}{3}\rho. \label{ModFRW}
\eea
The last equation gives  the entropy-corrected Friedmann equation of the FRW universe by considering
gravity as an entropic force. Notice that the Planck-scale correction terms include higher powers of $H$ suppressed by the Planck scale, the only physical scale we have.

We then derive the Raychaudhuri equation which corresponds to Eq.(\ref{ModFRW}).
We need to find the relation between $\dot{H}$ and $H$. Since we have $H=\dot{a}/a$, then
\be
\dot{H}= -H^2+\frac{\ddot{a}}{a} \la{dotH}.
\ee
Substituting Eq.(\ref{acceleration3}) and Eq.(\ref{ModFRW}) into Eq.(\ref{dotH}) and using $\omega=p/\rho$, we get:
\bea
\dot{H}&=& -\frac{3(1+\omega)}{2} H^2- \frac{1+3\omega}{2}\frac{k}{a^2}\nn\\
&+&\frac{\delta \ell_p^2}{18 \pi} \frac{(1+3\omega)^2}{2(1+\omega)} \le(H^2+\frac{k}{a^2}\ri)^2 \nn\\
&+&\frac{\delta \ell_p^2}{4 \pi} \frac{1+3\omega}{3(1+\omega)} \le(H^2+\frac{k}{a^2}\ri)^2
 \ln{\le(\frac{\ell_p^2}{4 \pi}\le(H^2+\frac{k}{a^2}\ri)\ri)} \la{dotH3}.
\eea

\subsection{Solutions of the modified Raychaudhuri equation}

In this section we solve the modified Raychaudhuri equation (\ref{dotH3}) for the flat case ($k=0$), using a perturbation method. For $k=0$, equation (\ref{dotH3}) has the form
\bea
\dot{H}= -\frac{3(1+\omega)}{2} H^2+\frac{\delta \ell_p^2}{36 \pi} \frac{(1+3\omega)^2}{(1+\omega)}\,H^4\,\le[
1+\frac{6}{1+3\omega} \ln{\le( {\ell_p H \over \sqrt{4 \pi}}\ri)} \ri]\la{dotH3-0}.
\eea
It is interesting to observe that this equation has a fixed point at $H\sim \ell_p^{-1}$ (see for example \cite{adel_1}), i.e. a de Sitter space, which smoothes the big bang singularity. As one might expect, the approximation used so far, $H \ell_p <1$, breaks down around the fixed point and we need an exact treatment to describe the solution near this point. Therefore, one can only trust a perturbative solution where $H \ell_p <1$ which we write as
\be
H(t)=H_0+ \lambda H_1
\ee
where $\lambda=\delta \ell_p^2$ and $H_0(t)=1/(\gamma\,t+C_1)$ is the solution of Eq.(\ref{dotH}) when $\lambda=0$.

Solving Eq.(\ref{dotH}) up to a leading order in $\lambda$ and imposing initial conditions we get

\bea
H(t)&=&\frac{H_0}{\le(\gamma H_0(t-t_0)+1\ri)}\nonumber\\
&+& \lambda \,
{\beta \,  {H_0}^3 \xi \over  \gamma\le( \gamma H_0(t-t_0)+1\ri)^2 } \Bigg[1+{(\sigma/\xi)\, \ln{(\gamma H_0(t-t_0)+1)}-1 \over \le( \gamma H_0(t-t_0)+1\ri)}\Bigg], \la{h1}
\eea
where $H_0$ is the initial Hubble parameter at time $t_0$, $\gamma = 3/2(1+\omega)$, $\beta = {(1+3\omega)^2/[36 \pi (1+\omega)]}$,  $\sigma=6/(1+3\omega)$ and $\xi=\sigma \ln{\le( {\ell_p H_0 / \sqrt{4 \pi}}\ri)}+1-\sigma$.

The scale factor can be calculated using the relation $H(t)=\dot{a}/a$, and it is given to the first
order of $\lambda$ as follows:

\bea
{a(t)}&=&{a_0}\le( \gamma H_0(t-t_0)+1\ri)^{1/\gamma}\Bigg[1-\lambda\, {\beta \, H_0^2 \eta \over 2 \gamma^2 \le( \gamma H_0(t-t_0)+1\ri)}\Bigg(1 \nonumber\\
&+&{{\sigma/\eta \ln{\le( \gamma H_0(t-t_0)+1\ri)} -1\over \le( \gamma H_0(t-t_0)+1\ri)}}\Bigg) \Bigg],
\eea
where $\eta=\sigma \ln{\le( {\ell_p H_0 / \sqrt{4 \pi}}\ri)}+1-3/2\,\sigma$.

Although, it is not clear from the approximation we have here if these corrections are going to resolve big bang singularities or not, a possible application of these corrections is early cosmology, in particular inflation models where physical scales are only a few orders of magnitude less than Planck scale.

\section{GUP linear and quadratic in $\D p$ and BH thermodynamics}

In this section, we will repeat the same analysis that was done in section 2, but with GUP linear
and quadratic in $\D p$.~It has been found in \cite{Ali:2012mt0,Ali:2012mt1}, that the inequality which
would correspond to Eq. (\ref{comm01}) can be written as follows:

\be \D x \D p \geq \frac{\hbar}{2} \le[1- \a_0 ~\ell_p~
\le(\frac{4}{3}\ri)~\sqrt{\mu}~~ \frac{\D p}{\hbar}+ ~2~
(1+\mu)~ \a_0^2 ~\ell_p^2 ~ \frac{\D p^2}{\hbar^2} \ri]\,.
\la{ineqII} \ee
~Solving it  as a quadratic equation in $\D p$ results in
\be \frac{\D p}{\hbar}\geq\frac{2 \D
x+\a_0
~\ell_p~\le(\frac{4}{3}~\sqrt{\mu}~\ri)}{4~(1+\mu)~\a_0^2~\ell_{p}^2}\le(1-
\sqrt{1-\frac{8~(1+\mu)~\a_0^2\ell_{p}^2} {\le(2 \D x+\a_0
\ell_p\le(\frac{4}{3}\ri) ~\sqrt{\mu}~\ri)^2}}\ri)
\,.\la{gupso} \ee
The negatively-signed solution is considered as the one that refers to the standard uncertainty relation as $\ell_p/\D x \rightarrow 0$. Using the Taylor expansion, we find that
\be
\Delta p \geq \frac{1}{\Delta x} \le(1- \frac{2}{3}\a_0 \ell_p \sqrt{\mu} \frac{1}{\Delta x} \ri).
\ee
We repeat the same analysis of Sec. (\ref{sec:GUP-QUAD}), and we get
\be
\Delta A_{min}= \lambda  \ell_{p}^2 \le[1- \frac{2}{3}\, \a_0\, \ell_p\, \sqrt{\frac{ \mu \, \pi}{A}}\ri],
\ee
Which leads to the modified entropy as
\be
S=\frac{A}{4\, \ell_p^2} + \frac{2}{3}\, \a_0\,  \sqrt{\pi\, \mu\, \frac{A}{4\, \ell_p^2}}. \la{correctENTROPY}
\ee
We find that the entropy is directly related to the area and is modified when applying GUP-approach. The temperature of the black hole is
\bea
T&=& \frac{\kappa}{8 \pi} \frac{dA}{dS}
=\frac{\kappa}{8 \pi} \le[1- \frac{2}{3}\, \a_0\, \ell_p\, \sqrt{\mu\, \frac{\pi}{A}} \ri].
\eea
We find that the temperature is not only proportional to the surface gravity but
also it depends on the black hole's area.

\subsection{Modified Newton's law of gravity}

In this section we study the implications of the corrections
calculated for the entropy in  Eq.~(\ref{correctENTROPY})
and calculate how the number of bits of Eq.~(\ref{Nbits})
would be modified to identify new corrections
to Newton's law of gravitation. Using the corrected entropy
given in Eq. (\ref{correctENTROPY}), we find that the number of bits
should also be corrected as follows.

\be
N^{\prime\prime}= \frac{4 S}{k_B}= \frac{A}{\ell_{p}^2}+ \frac{4}{3}\, \a_0\, \sqrt{\mu\, \pi\, \frac{A}{\ell_{p}^2}}. \la{bits}
\ee
By substituting Eq. (\ref{bits}) into Eq. (\ref{energy}) and using Eq. (\ref{force}), we get
\be\label{eq:Ee}
E = 
      F\, c^2\, \le(\frac{r^2}{m\, G}+ \frac{\a\, \sqrt{\mu}\, r}{3\, m\, G}\ri).
\ee
It is apparent, that Eq. (\ref{eq:Ee}) implies a modification to Newton's law of gravitation \cite{Ali:2013ma};
\be
F= G \frac{M m}{r^2} \le(1- \frac{\a\, \sqrt{\mu}}{3\, r}\ri). \la{result}
\ee
This equation states that the modification in Newton's law of gravity seems to agree with the predictions of the Randall-Sundrum II model \cite{Randall:1999vf} which contains one uncompactified extra dimension and length scale $\Lambda_R$. The only difference is the sign. The modification in Newton's  gravitational potential on the brane \cite{potential} is given as
\bea
V_{RS} =\begin{cases} -G\frac{m M}{r} \le(1+\frac{4 \Lambda_R}{3 \pi r}\ri), &  r\ll\Lambda_R\\
& \\
-G\frac{m M}{r} \le(1+\frac{2 \Lambda_R}{3 r^2}\ri), & r\gg\Lambda_R
\end{cases}, \la{VRS}
\eea
where $r$ and $\Lambda_R$ are the radius and the characteristic
length scale respectively.
The result, Eq. (\ref{result}), agrees with Eq. (\ref{VRS})
, albeit with the opposite sign when  $r\ll\Lambda_R$. This result says that
$\alpha \sim \Lambda_R$  which helps to set a new upper bound on the value of the parameter $\alpha$.
This means that the proposed GUP-approach \cite{advplb,Das:2010zf} is
apparently able to predict the same physics as Randall-Sundrum
II model.
In recent gravitational experiments,
it is found that the Newtonian gravitational force, the $1/r^2$-law,
seems to be maintained up to $\sim0.13-0.16~$mm \cite{gt20,gt21}. However,
it is unknown whether this law is violated or not at sub-$\mu$m range.
Further implications of these modifications have been discussed in
\cite{Buisseret:2007qd} which could be the same for the GUP
modification. This similarity between the GUP and
the extra dimensions of the Randall-Sundrum II model would lead to new bounds on the GUP parameter $\alpha$ with respect to
the extra dimension length scale $\Lambda_R$. Further investigations are needed.

\subsection{Modified Friedmann and Raychaudhuri Equations}

In this subsection, we calculate the quantum gravity corrections predicted by the GUP of Eq. (\ref{comm01}).
we repeat the same analysis that was done in subsection (\ref{sec:GUP-Friedmann}), but this time
using the modified Newton's law of gravity of Eq. (\ref{result}).

After making the same analysis of subsection (\ref{sec:GUP-Friedmann}), we get the modified  acceleration equation for the dynamic evolution of the FRW universe as follows:
\be
\frac{\ddot{a}}{a}=-\frac{4 ~\pi G}{3} (\rho+3 p) \le[1-\frac{\a~ \sqrt{\mu}}{3~ \tilde{r}}\ri]. \label{acceleration4}
\ee
Using the continuity equation of Eq.(\ref{Continuity}) in Eq.(\ref{acceleration3}) and multiplying both sides with $(a~\dot{a})$, we get after making integration the following equation:
\be
\dot{a}^2+k= \frac{8\pi G}{3}\rho a^2\le[1-\frac{\a \sqrt{\mu}}{3} \frac{1}{\rho~a~\tilde{r}} \int \frac{d(\rho ~a^2)}{a}\ri].
\ee
By substituting Eq.(\ref{rho}) into Eq.(\ref{adot}) and integrating, we get
\be
H^2+\frac{k}{a^2}=\frac{8~ \pi~ G}{3} \rho \le[1-\frac{\a \sqrt{\mu}}{3} \frac{1+3\omega}{2+3\omega} \frac{1}{\tilde{r}}\ri].
\ee
Using the relation  $\tilde{r}=a ~r=\frac{1}{\sqrt{H^2+k/a^2}}$, we get:
\be
H^2+\frac{k}{a^2}=\frac{8~ \pi~ G}{3}  \rho ~\le[1-\frac{\a \sqrt{\mu}}{3} \le(\frac{1+3\omega}{2+3\omega}\ri) \sqrt{H^2+\frac{k}{a^2}}\ri],
\ee
which can be written as:
\be
\le(H^2+\frac{k}{a^2}\ri) \le[1+\frac{\a \sqrt{\mu}}{3} \le(\frac{1+3\omega}{2+3\omega}\ri) \sqrt{H^2+\frac{k}{a^2}}\ri]=\frac{8~ \pi~ G}{3}  \rho.~ \label{ModFRW1}
\ee
We then derive the Raychaudhuri equation which corresponds to Eq.(\ref{ModFRW1})
in which we find the relation between $\dot{H}$ and $H$.
\be
\dot{H}= -H^2+\frac{\ddot{a}}{a}. \la{dotH1}
\ee
Substituting Eq.(\ref{acceleration4}) and Eq.(\ref{ModFRW1}) into Eq.(\ref{dotH1}), and using $\omega=p/\rho$, we get:
\bea
\dot{H}&=& -\frac{3(1+\omega)}{2}H^2-\frac{1+3\omega}{2}\frac{k}{a^2}\nn\\
&+&\frac{\a \sqrt{\mu}}{3}\frac{1+3\omega}{2(2+3\omega)}\le(H^2+\frac{k}{a^2}\ri)^{3/2}.\la{dotH4}
\eea

\subsection{Solutions of modified Raychaudhuri equation}

Here we discuss solutions of the modified Raychaudhuri equation (\ref{dotH4-0}) for the flat case, $k=0$. In this case, equation (\ref{dotH4}) has the form
\bea
\dot{H}= -\frac{3(1+\omega)}{2}H^2 +\frac{\a \sqrt{\mu}}{3}\frac{1+3\omega}{2(2+3\omega)} H^3\la{dotH4-0}.
\eea
Similar to the quadratic GUP case, the above equation has a fixed point at $H\sim \ell_p^{-1}$. The above equation was studied by Murphy \cite{murphy}, where an exact solution was presented. Recently, the above equation was studied in \cite{adel_1} as an example to show that pressure properties, such as asymptotic behavior and fixed points can be used to qualitatively describe the entire behavior of a FLRW solution. Similar to the quadratic case, the approximation i.e., $H \ell_p <1$, breaks down around the fixed point and one should have an exact treatment to describe the solution near this point. Here we are interested in a perturbative solution, which can written as
\be
H(t)=H_0+ \epsilon H_1
\ee
where $\epsilon=\frac{\a \sqrt{\mu}}{6}$ and $H_0(t)=1/(\gamma\,t+C_1)$ is the solution of Eq.(\ref{dotH}) as we set $\epsilon=0$.\\

Solving Eq.(\ref{dotH}) up to a leading order in $\epsilon$ and imposing initial conditions we get
\bea
H(t)=\frac{H_0}{\le(\gamma H_0(t-t_0)+1\ri)}+\epsilon \,
{ \beta' \,  {H_0}^2 \ln{\le(\gamma H_0(t-t_0)+1\ri)} \over \gamma \le( \gamma H_0(t-t_0)+1\ri)^2 } \la{h2}
\eea
where $H_0$ is the Hubble parameter at initial time $t_0$, $\gamma = 3/2(1+\omega)$, $\beta' = {(1+3\omega)/(2+3\omega)}$.

The scale factor can be calculated using the relation $H(t)=\dot{a}/a$, and it is given to the first
order of $\epsilon$ as follows:
\bea
{a(t)}={a_0}\,\le(\gamma H_0(t-t_0)+1\ri)^{1/\gamma}\Bigg[1-\epsilon\, {\beta' \, H_0 \over \gamma^2}{\ln{\le(\gamma H_0(t-t_0)+1\ri)} \over \le(\gamma H_0(t-t_0)+1\ri)}\Bigg]
\eea

\section{Conclusions}

Verlinde has extended the validity of the holographic
principle to assume a new origin of gravity as an entropic force.
We have used this extension to calculate corrections to Newton's law of gravitation
as well as the Friedmann equations arising from the generalized uncertainty principle suggested
by different approaches to quantum gravity such as string theory and black hole physics. We followed the following procedure:
\emph{modified theory of gravity $\rightarrow$ modified black hole entropy$\rightarrow$
modified holographic surface entropy $\rightarrow$� Newton's law corrections$\rightarrow$ modified Friedmann equations}.

We found that the corrections for Newton's law of gravity agree with the brane world corrections.
This suggests that GUP and brane world may have very similar gravity at low-energy. Besides, we note that the derived correction terms for the Friedmann
equation vanishes rapidly with increasing apparent horizon radius, as expected. This means that the corrections become relevant
at the early universe, in particular, with the inflationary models where the physical
scales are few orders of magnitude less than the Planck scale. When the
universe becomes large, these corrections can be ignored and the modified  Friedmann equation reduces to the standard Friedman
equation. We can understand that  when $a(t)$ is large, it is difficult to excite
these modes and hence, the low-energy modes dominates the entropy.~But at the early universe, a large number
of excited modes can contribute causing a modification to the area law leading to the modified Friedmann equations.

But could we observe the impact of these corrections on the early universe? Since these corrections modify the standard FRW cosmology, especially in early times, it is expected that they have some consequences on inflation. One of the interesting results reported in the Planck 2013 data \cite{Ade:2013rta} is that exact scale-invariance of the scalar power spectrum has been ruled out by more than $5\sigma$. Meaning that, the early universe tiny quantum fluctuations, which eventually cause the formation of galaxies, not only depend on the mode wave number k, but also on some physical scale. This shows that scalar power spectrum and other inflation parameters could depend on physical scale. The energy scale of inflation models has to be around Grand Unified Theories (GUT) scale or larger, therefore, this cutoff could be the Planck scale.

There are two distinct possibilities to introduce the Planck scale to modify the standard inflation scenario. The first possibility is to use it as a momentum-cutoff in the quantum field theory of the inflaton field (see for example \cite{danielsson} and references there in). The second possibility is to modify general relativity, which will lead to a modified Friedmann equation. The latter framework is considered in several interesting inflationary models, such as Brane-world, and $f(R)$ inflationary models. Consequences of modifying Friedmann's equation on inflation have been discussed in literature, for example see \cite{delcampo} and references there in.

Our framework lies in the second class which modifies the Hubble parameter $H$ as a function of time compared to that of the standard FRW cosmology, as expressed in equations (\ref{h1}) and (\ref{h2}). Since the slow-rolling parameters  $\epsilon$ and $\eta$ depend on $H$, changes in $H$ will affect them and the scalar spectral index $n_s$, which will have an impact on observations. The coming generations of CMB precise observations will be able to measure the inflation parameters with high accuracy, therefore enabling different inflationary models to be distinguished. In the future, it would be interesting to apply our approach to investigations of these modified Friedmann's equations on specific inflationary models as well as the reheating phase of the universe. We hope to report on these issues soon.

\section*{Acknowledgments}
The research of AFA is supported by Benha University (www.bu.edu.eg) and CTP in Zewail City.~AFA would like to thank ICTP, Trieste for the kind hospitality where the present work was finished. The authors thank the anonymous referees for useful comments and suggestions.


\end{document}